\documentclass[5p]{elsarticle}
\usepackage[english]{babel}
\usepackage{graphicx}
\usepackage{dcolumn}
\usepackage{bm}
\usepackage[latin1]{inputenc}
\usepackage{graphicx}
\usepackage{amsmath}
\usepackage{amssymb}
\usepackage{pifont}
\usepackage{epstopdf}
\usepackage{color}

\DeclareMathOperator{\arsinh}{arsinh}

\graphicspath{{figs/}}

\begin{document}

\title{Slippery and mobile hydrophobic electrokinetics: from  single walls  to nanochannels}

\author{Olga I. Vinogradova\corref{cor1}}
\ead{oivinograd@yahoo.com}
\author{Elena F. Silkina}
\author{Evgeny S. Asmolov}
\cortext[cor1]{}

\address{Frumkin Institute of Physical Chemistry and Electrochemistry, Russian Academy of Sciences, 31 Leninsky Prospect, 119071 Moscow, Russia}

\begin{abstract}
We discuss how the wettability of solid walls impacts electrokinetic properties, from large systems to a nanoscale. We show in particular how could the  hydrophobic slippage, coupled to confinement effects,  be exploited to induce novel electrokinetic properties, such as a  salt-dependent giant amplification of zeta potential and conductivity, and a much more efficient  energy conversion. However, the impact of slippage is dramatically reduced if some surface charges migrate along the hydrophobic wall under an applied field.
\end{abstract}

\maketitle

\section{Introduction}

\begin{figure}[t]
	\begin{center}
		\includegraphics[width=0.98\columnwidth , trim=0.cm 0. 0.0cm
		0.,clip=false]{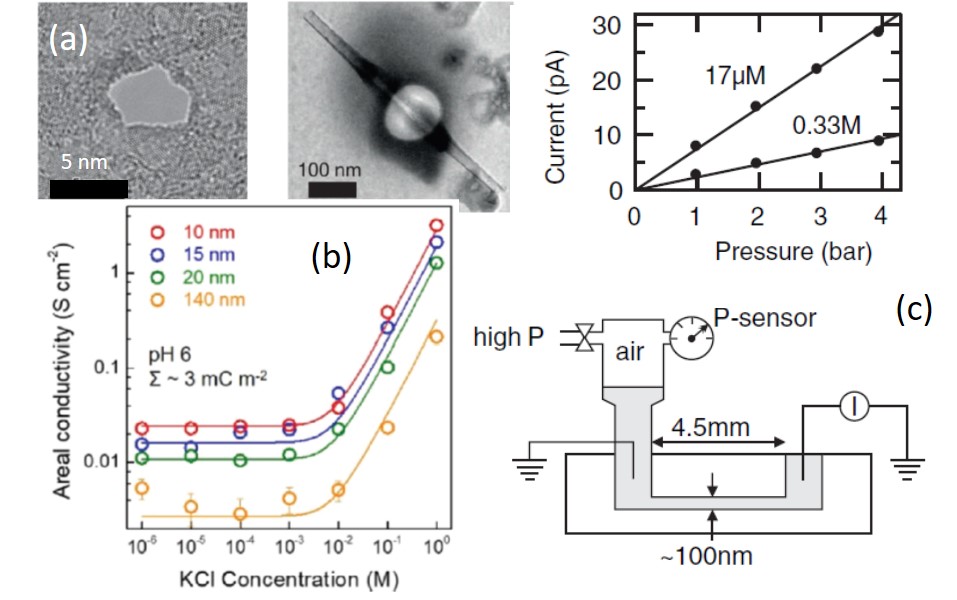}
	\end{center}
	\caption{(a) Nanopore in single layer  MoS$_2$~\cite{feng.j:2016} and boron nitride nanotube inserted into a SiN membrane~\cite{siria.a:2013}, (b) Salt dependence of conductivity for membranes made of holey-graphene-like network of different thicknesses \cite{wang.h:2020}, (c) Streaming current as a function of pressure for a 140 nm thick fused silica slit at low and high salt concentrations and schematics of this experiment~\cite{vanderHeyden.fhj:2005}.
}
	\label{fig:collage}
\end{figure}

Controlling the wettability of solid materials is a key issue in surface engineering, and wetting phenomena have been studied
for at least two centuries. In many common situations the equilibrium contact angle lies between $0$ and $90^{\circ}$ (i.e.,
a hydrophilic case), but some solids can have a contact angle greater than $90^{\circ}$ (a hydrophobic case). This can generate some special properties of practical
interest, based on water repellency, such as, for example, an extremely long-ranged hydrophobic attraction. Over past few decades, the cutting edge in the
research on smooth hydrophobic surfaces has shifted from equilibrium wetting and surface forces towards properties that can
impact the dynamics of liquids due to hydrophobic (hydrodynamic) slippage.
In the last years research on hydrophobic slippage is rapidly advanced being strongly motivated by potential applications, such as drag reduction and more. These  hydrodynamic studies recently raised a question of a possible influence of hydrophobic slippage on the electrokinetic transport phenomena in micro- and nanofluidic channels, which include an electro-osmotic flow in response to an applied electric field, a conductance, emerging due to this flow and a migration of ions, and also a streaming current generated by pressure gradient (see Fig.\ref{fig:collage}).
This was also  motivated  by an awareness of many experimental  puzzles, such as an enormous conductivity of dilute electrolyte solutions confined in nanochannels~\cite{stein.d:2004,schoch.rb:2005} enhanced by hydrophobization of the walls~\cite{balme.s:2015}, the saturation of the conductivity of nanometric foam films~\cite{bonhomme.o:2017}, and more.

In this article we review recent developments in the theory of electrokinetic transport near a hydrophobic wall and in hydrophobic channels by concentrating on the fundamental understanding and expectations. Our main focus will be on channels of micro- and nanometric size, where transport phenomena can be very strongly modified by tuning the  electrostatic and wetting properties of the confining surfaces. We attempt to give the flavour of some of the recent  work  in this field, but our review is not intended  to be comprehensive. Emphasis is placed to a continuum description of electrokinetic phenomena; the choice reflects the authors own interests. Thus, our discussion applies for channels down to a few nanometers and we limit ourselves by 1:1 electrolyte solutions of the concentration range from $10^{-6}$ to $10^{-1}$ mol/l, which provides a very accurate description of the ionic distributions within the Poisson-Boltzmann theory~\cite{poon.w:2006}. Correlations and various nonidealities,  such as hydrated ion volume effects~\cite{zhu.h:2019}, dielectric mismatch~\cite{bonthuis.dj:2012}, dispersion forces between ions~\cite{ninham.bw:1997},  ion-specifity~\cite{cao.q:2018}, which would be important for channels of one or two nanometers or in molecular-scale confinement, warrant a separate discussion. We, therefore, refer the readers to the recent  reviews on sub-continuum electrokinetic effects~\cite{kavokin.n:2021} and relevant simulation techniques~\cite{pagonabarraga.i:2010,hartkamp.r:2018,gubbiotti.a:2022}.

\section{Electrokinetic transport: electro-osmosis, conductivity currents
and beyond}

\begin{figure}[t]
	\begin{center}
		\includegraphics[width=0.7\columnwidth , trim=0.cm 0. 0.0cm
		0.,clip=false]{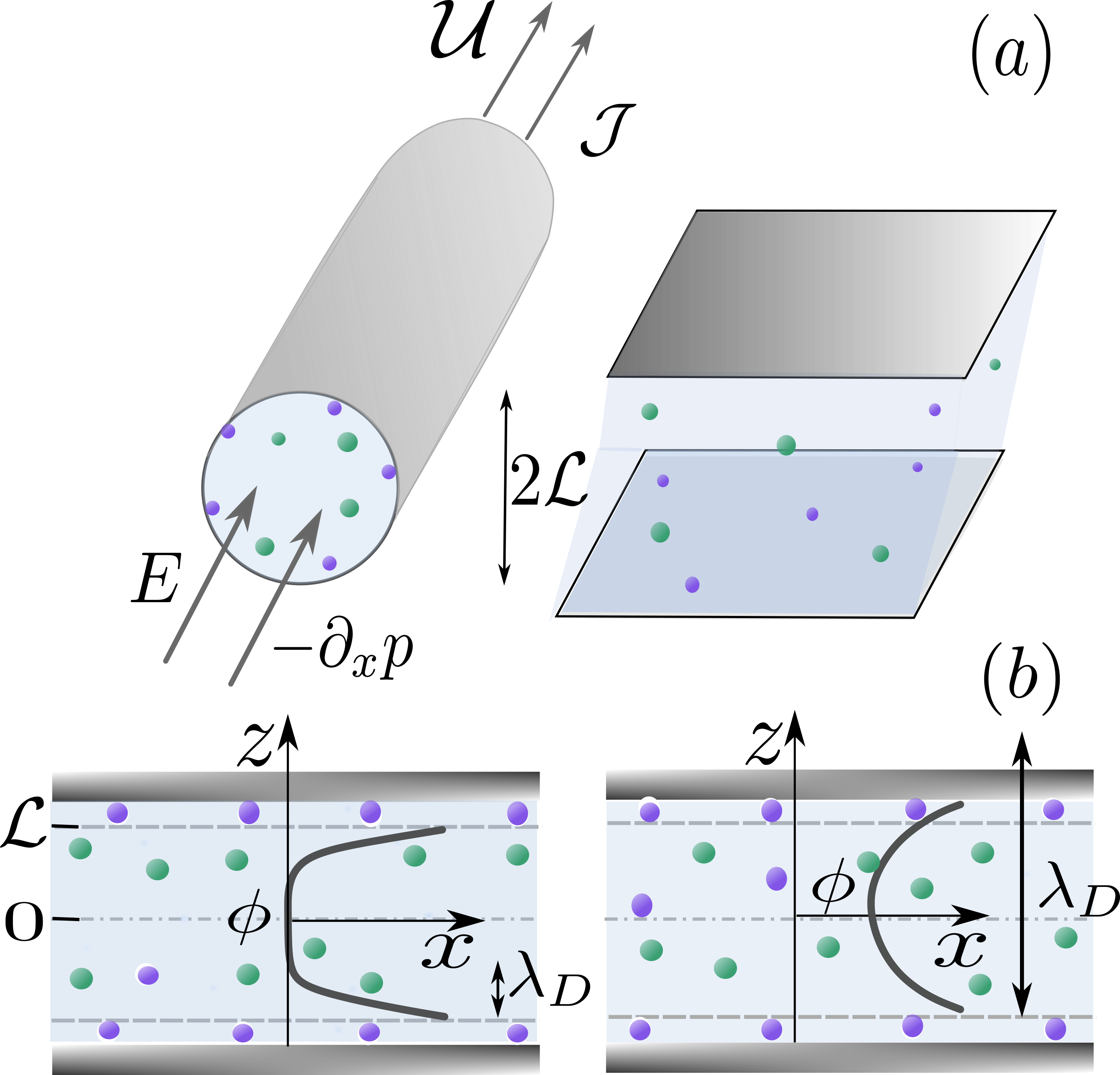}
	\end{center}
	\caption{(a) Sketch of the nanotube and slit. An applied electric field $E$ and/or a pressure gradient $\partial_{x}p$ generate a mean fluid velocity $\mathcal{U}$ and electric current of the mean density $\mathcal{J}$.  (b) Profiles of a dimensionless electrostatic potential $\phi(z)$ in thick and thin channels.}
	\label{fig:sketch}
\end{figure}

One imagines a bulk electrolyte solution of a dynamic viscosity $\eta $
and permittivity $\varepsilon $ in contact with a symmetric channel. In the general case, the channel subject to a pressure gradient $\partial _{x}p$ and an electric field $E$ in the $x$ direction as sketched in Fig.~\ref{fig:sketch}.
The axis $z$ is defined normal to the surfaces of potential $\Phi_s$ and charge density $\sigma$ (without loss of generality, the
surface charges are taken here as cations) located at $z=\mathcal{L}$. Although channels of variable thickness have been considered as a model for certain pores~\cite{gravelle.s:2013,malgaretti.p:2019}, two main geometries are still favoured: slits and cylinders. In the former the electrolyte is confined between two parallel walls of (infinite) area, separated by a finite distance $H$. For such a channel it is enough to consider $\mathcal{L}= H/2$ because of the $z \leftrightarrow - z$ symmetry. For the cylinder geometry the interior radius $\mathcal{L} = R$ is finite but the length is infinite.
Let the bulk reservoir represent a 1:1 salt solution of number density of ions $n_{\infty}[\rm{m^{-3}}]$. Clearly, by analysing the experimental data it is more convenient to use the concentration $c_{\infty}[\rm{mol/l}]$, which is related to a number density
as $n_{\infty} = N_A \times 10^3 \times c_{\infty}$, where $N_A$ is Avogadro's number.
Ions are assumed to obey the Boltzmann distribution $n_{\pm }(z)=n_{\infty}\exp (\mp \phi (z))$, where $\phi (z)=e\Phi(z)/(k_{B}T)$ is the dimensionless local electrostatic
potential, $e$ is the elementary positive charge, $k_{B}$ is the Boltzmann
constant, $T$ is a temperature of the system, and the upper (lower) sign
corresponds to the positive (negative) ions.

An electrolyte solution builds up a so-called electrostatic diffuse layer (EDL) close to the channel walls, where the surface charge is balanced by the cloud of counterions. In thick channels EDLs do not overlap and at their central part contains the (electro-neutral) bulk electrolyte, so $\Phi(0) = 0$. However, when the channel is sufficiently thin, the EDLs overlap.   There exists no  bulk electrolyte solution inside and $\Phi(0) \neq 0$ - see Fig.~\ref{fig:sketch}.

\begin{figure}[t]
\begin{center}
\includegraphics[width=0.45\columnwidth , trim=0.cm 0. 0.0cm
0.,clip=false]{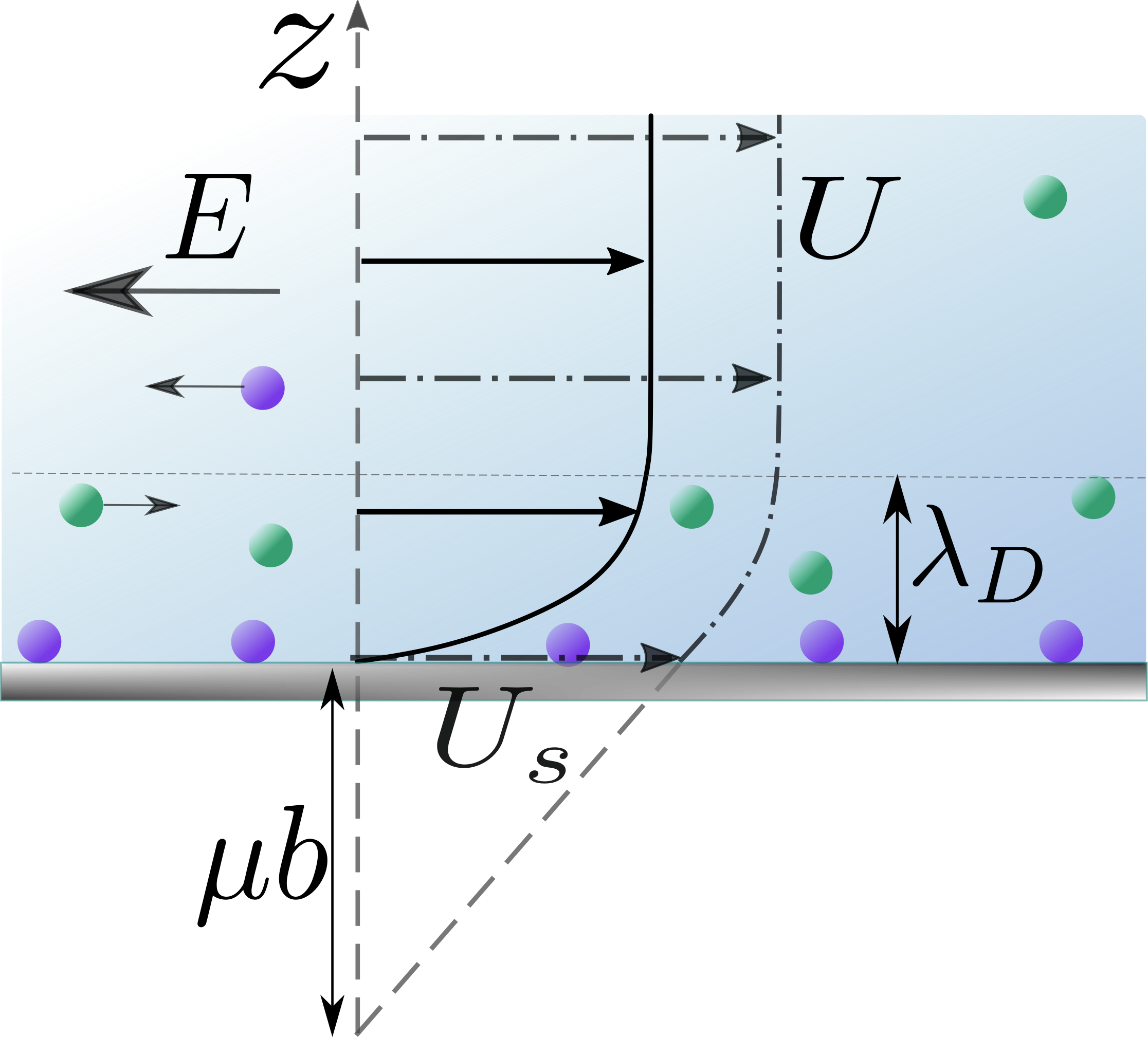}
\end{center}
\caption{Sketch of the  electroosmotic velocity profiles $U(z)$ near  positively charged hydrophilic (a solid curve) and hydrophobic (a dash-dotted curve) single walls of a potential $\Phi_s$. A finite slip velocity $U_s$ at the hydrophobic wall causes the difference between $\Phi_s$ and $Z$. }
\label{fig:smol}
\end{figure}

When an  electric field $E$ is applied tangent to a charged wall, an electro-osmotic flow of a velocity $U(z)$ is induced. The electroosmosis takes its  origin in the  EDL, where a tangential electric
field $E$ generates a force that, in turns, sets the fluid in motion. The successful understanding of electro-osmosis is due to Smoluchowski~\cite{smoluchowski.m:1921}, who argued that the velocity of a plug flow in the bulk is given by
\begin{equation}\label{eq:smoluchowsky}
  U = - \dfrac{\varepsilon E}{4 \pi \eta } \Phi_s
\end{equation}
Eq.\eqref{eq:smoluchowsky} implies that in the case of the positively charged surface, the direction of fluid flow is opposite to that of the electric field (see Fig.~\ref{fig:smol}). Note that Smoluchowski  considered the single wall ($\mathcal{L} \to \infty$) and postulated the no-slip boundary condition. Only if so, the electrokinetic  potential, later termed zeta potential  $Z$, that   as a matter of fact should appear in \eqref{eq:smoluchowsky} via the Stokes
equation, coincides with the surface potential $\Phi_s$.  However, the conclusion that $Z = \Phi_s$, which became a dogma in colloidal science and long time invoked in the interpretation of the electrokinetic data, is by no means obvious for a sufficiently thin channel and/or for a situation where the no-slip boundary condition is violated.

An applied field also generates an electric (so-called conductivity) current that is both due to their convective transfer by an emerging electroosmotic flow and a migration of ions relative to a solvent. The bulk conductivity can be found as $K_{\infty}  = 2 e n_{\infty} m$, where $m$ is the mobility of ions. In the simplest case, the electric force $e E$, causing the migration of ions, is balanced by the Stokes force, which for ions of hydrodynamic radius $\mathcal{R}$ (of the order of a few tenths of nm) gives $m = e/ (6\pi \eta \mathcal{R})$. Consequently,
\begin{equation}
	K_{\infty }=\frac{e^{2} n_{\infty}}{3 \pi \eta \mathcal{R}}
	 \label{eq:ic0_thick}
	\end{equation}
Clearly, if $\mathcal{L}$ is large enough, an average ionic conductivity of a confined solution turns to $K_{\infty }$.

 Similarly, the pressure-driven flow induces a convective transfer (but no migration) of ions and thus generates a current, referred to as streaming current. Since the streaming current can only be generated within the EDLs, but not in the electro-neutral bulk electrolyte, its depth-averaged value becomes negligible in the large systems.

\section{Spectrum of electrostatic lengths}

There are a number of electrostatic length scales that control electrokinetic  transport of water and ions. These length scales
are related to properties of either electrolyte solution or surfaces, or both. Specific micro- and nanofluidic phenomena will
show up when the channel size $\mathcal{L}$ becomes of the order of or smaller than these characteristic lengths. Another advantage of using a dimensionless potential and characteristic  lengths is that the results become independent on the choice of  any specific system of electrostatic units.

\begin{itemize}
  \item The so-called Bjerrum length is a length for which the thermal energy is equal to the Coulombic
energy between two unit charges
\begin{equation}\label{eq:bjerrum}
 \ell _{B}=\dfrac{e^2}{\varepsilon k_B T}
 \end{equation}
Defined by such a way the Bjerrum length is a property of solvent and does not depend on the electrolyte concentration.  For water at a room temperature $\ell_{B} \simeq 0.7$ nm. We remark that $\ell _{B}/\mathcal{R} =O(1)$.

  \item The extension of the EDLs is defined by the Debye (screening) length
\begin{equation}\label{eq:DH_length}
      \lambda_{D}=\left( 8\pi \ell _{B} n_{\infty}\right) ^{-1/2}
     \end{equation}
of a bulk solution. For a monovalent salt in water at room temperature $\lambda_D [\rm{nm}] \simeq \dfrac{0.3 [\rm{nm}]}{\sqrt{c_{\infty}} [\rm{mol/l}]}$
Therefore, by increasing $c_{\infty}$ from $10^{-6}$ to $10^{-1}$ mol/l, we reduce the screening length ca. from  300 down to 1 nm. The channel of $\mathcal{L} \gg \lambda_D$ is traditionally referred to as thick and of $\mathcal{L} \ll \lambda_D$ is termed thin.

  \item The Gouy-Chapman length is inversely proportional to the surface charge density $\sigma$
\begin{equation}\label{eq:LGC}
  \ell _{GC}=\dfrac{e}{2\pi \sigma \ell _{B}}
\end{equation}
It is often convenient to define $\ell _{GC}$ of the same sign
with $\sigma $,  although some researchers use only positive definite $\ell _{GC}$.
For monovalent ions in water at room temperature $\ell _{GC}[\mathrm{nm}] \simeq \dfrac{36 [\mathrm{nm}]}{\sigma [\mathrm{mC/m}^2]}$. A typical (high) surface charge density $\sigma \simeq 36 $ mC/m$^2$ gives $\ell _{GC} \simeq 1$ nm, but small $\sigma$ provides much larger $\ell _{GC}$. The important ratio is $\lambda_D/\ell _{GC} \propto \sigma n_{\infty}^{-1/2}$, which reflects the \emph{effective surface charge}. The surfaces are referred to as weakly charged when $\lambda_D/\ell _{GC} \leq 1$, and to as strongly charged if $\lambda_D/\ell _{GC} \gg 1$. Say, a surface of $\ell _{GC} = 3$ nm is weakly charged when $c_{\infty} \geq 10^{-2}$ mol/l and strongly charged when $c_{\infty} \leq 10^{-4}$ mol/l.

  \item The Dukhin length is defined as
\begin{equation}  \label{eq:dukhin_length}
\ell_{Du} =  \frac{\lambda_D^2}{\ell_{GC}} =  \dfrac{\sigma}{4 e n_{\infty}}
\end{equation}
For a surface of $\sigma \simeq 18 $ mC/m$^2$ on increasing $c_{\infty}$ from $10^{-6}$ to $10^{-1}$ mol/l, one can reduce $\ell _{Du}$ from ca. 45 $\mu$m down to 0.5 nm. In very dilute solutions, therefore, $\ell_{Du}$ can be much larger than any conceivable Debye length.
\end{itemize}

\section{Governing equations and boundary conditions}\label{sec:GEBC}

The flow inside the channel satisfies the \emph{linear} Stokes equation with an electrostatic body force
\begin{equation}  \label{eq:Stokes}
\eta \nabla^2 U=\partial _{x}p+\dfrac{eE}{4\pi \ell _{B}}\nabla^2 \phi,
\end{equation}
where $U$ is the fluid velocity. One can also define a dimensionless velocity as $u = \dfrac{4\pi \eta
\ell _{B}}{e E} U$.

In steady state $\phi(z)$ is independent on the
fluid flow and satisfies the \emph{nonlinear} Poisson-Boltzmann
equation (NLPB)~\cite{andelman.d:2006book, herrero.c:2021}:
\begin{equation}  \label{eq:NLPB}
\nabla^2 \phi =\lambda _{D}^{-2}\sinh \phi,
\end{equation}
which can be linearized only if $\phi_s \leq 1$ (or $\Phi_s \leq 25$ mV).

When modeling electrokinetics in nanochannels and nanopores, special care should be given to the selection of boundary conditions. Clearly, the symmetry conditions $\partial _{z} u|_{z=0} = \partial _{z} \phi|_{z=0} = 0$ are always hold, but the conditions at the walls might be different depending on their material.

\subsection{Electrostatic boundary conditions and the contact theorem}\label{sec:BCCT}

To solve NLPB it is convenient to assume either a constant surface charge density (insulators)
\begin{equation}
\phi^{\prime} |_{z=\mathcal{L}}=\dfrac{2}{\ell _{GC}},  \label{eq:bc_CC}
\end{equation}
or a constant surface potential (conductors)
\begin{equation}
\phi |_{z=\mathcal{L}}=\phi _{s}  \label{eq:bc_CP}
\end{equation}
These situations, referred below to as CC and CP cases, provide rigorous  bounds on any solutions obtained by imposing a so-called charge regulation (CR) - another commonly used boundary condition.

The first integration of the NLPB equation leads to a relation between $\ell _{GC}$ and $\phi_s$, known as the \emph{contact theorem}. Once this relation is found, the NLPB solution obtained for the CC case can be immediately transformed into a solution for the CP case and vice versa.

Throughout this article we present analytic  results for the following two experimentally relevant modes:
\begin{itemize}
  \item In the limit of the thick channel, $\mathcal{L} /\lambda_D \gg 1$, the contact theorem coincides with the (exact) Grahame
equation for a single wall~\cite{israelachvili.jn:2011}
 \begin{equation}\label{eq:pot-charge_hs}
\phi _{s} = 2\arsinh\left(\frac{\lambda _{D}}{\ell _{GC}}\right)
\end{equation}
For weakly charged surfaces, $\lambda _{D}/\ell _{GC} \leq 1$, this equation can be linearized, so is the NLPB. In this case, $\phi_s \propto c_{\infty}^{-1/2}$.
Does the Graham equation applies for a channel of $\mathcal{L} /\lambda_D = O(1)$?  The answer is yes, but only if surfaces are strongly charged~\cite{vinogradova.oi:2021}. Such channels are termed quasi-thick and are said to be in the \emph{thick channel mode}.
    \item In the so-called \emph{thin channel mode} defined below, the contact theorem reads~\cite{silkina.ef:2019}
\begin{equation}
\phi _{s}\simeq  \arsinh \left(\dfrac{2 \alpha \ell_{Du}}{\mathcal{L}}\right),
  \label{eq:pot-charge_ls}
\end{equation}
where $\alpha$ reflects the geometry being equal to  $1$ for a slit, and to $2$ for a cylinder. Clearly, the linearization of \eqref{eq:pot-charge_ls} and, consequently, of \eqref{eq:NLPB} cannot be justified  when $\ell_{Du}/\mathcal{L} \geq 1$, even provided the effective charge is small. In essence, the derivation of Eq.\eqref{eq:pot-charge_ls} does not require a thin channel limit, $\mathcal{L} /\lambda_D \ll 1$, but only imposes that $\phi_s > \mathcal{L}/\ell _{GC}$. This is
equivalent to saying that the effective charge should be below $\phi_s \lambda_D/\mathcal{L}$ or that
$\mathcal{L} /\lambda_D < \ell _{GC} \phi_s/\lambda_D$. This condition can be satisfied even if $\mathcal{L} /\lambda_D = O(1)$.
To fix the idea, set $\mathcal{L}=\lambda_D=10$ nm and $\alpha = 2$. Choose $\ell _{GC} = 5$ nm. For such a nanotube Eq.\eqref{eq:pot-charge_ls} gives $\phi_s \simeq 2.8$, which exceeds $\mathcal{L}/\ell _{GC} = 2$. Thus, this nanotube is quasi-thin, i.e. does fall into a thin channel mode, but one can easily verify that the nanotube of $\ell _{GC} = 1$ nm, does not.
\end{itemize}

The derived below electrokinetic equations can be used both for CC and CP cases. For a CC channel $\ell_{GC}$ does depend on neither $c_{\infty}$ nor $\mathcal{L}$, and $\ell_{Du} \propto c_{\infty}^{-1}$. The expressions for  $\ell_{Du}$ and $\ell_{GC}$ of a CP channel, depending on the mode, follow directly from Eqs.\eqref{eq:pot-charge_hs} or \eqref{eq:pot-charge_ls} and are summarised in Table~\ref{table:CPCC}.

\begin{table}
  \centering
    \renewcommand{\baselinestretch}{2}\normalsize
  \caption{The Dukhin and Gouy-Chapman lengths for a CP channel.}
        \label{table:CPCC}
  \begin{tabular}{|c|c|}
  \multicolumn{2}{c}{} \\
  \hline
   Thick channel mode & Thin channel mode \\
       \hline
    $ \ell_{Du} =  \lambda_D \sinh \dfrac{\phi _{s}}{2} \propto c_{\infty}^{-1/2}$  &  $\ell_{Du} \simeq \dfrac{\mathcal{L} \sinh \phi_s}{2\alpha} \propto c_{\infty}^{0}$  \\
    \hline
     $\ell_{GC} = \dfrac{\lambda_D}{\sinh \dfrac{\phi _{s}}{2}}\propto c_{\infty}^{-1/2}$  & $\ell_{GC} \simeq \dfrac{2\alpha \lambda_D^2}{\mathcal{L} \sinh \phi _{s}} \propto c_{\infty}^{-1}$ \\
                  \hline
        \end{tabular}
        \end{table}

Note that some approximate expressions for a conductance in thin nanopores have been also obtained by  either postulating the uniform potential across the pore (``Donnan equilibrium'')~\cite{biesheuvel.pm:2016,peters.pb:2016}, or for the ``counter-ions only'' case (so-called co-ion exclusion approximation)~\cite{uematsu.y:2018,green.y:2022}. To what extent and when can these primitive models be employed is discussed in a recent review~\cite{herrero.c:2021}.

\subsection{Hydrodynamic boundary conditions and the slip length}

The fluid flow at interfaces introduces an additional, hydrodynamic length scale of the problem. Namely, the so-called slip length, $b$, that is defined as
\begin{equation}
U|_{z=\mathcal{L}}= - b \partial _{z} U|_{z=\mathcal{L}}  \label{eq:bc_slip}
\end{equation}%
Equation~\eqref{eq:bc_slip} represents a boundary condition for a pressure driven flow. Physically, the (scalar) slip length characterizes the friction of the fluid at
the interface, and large $b$, indicating low friction, is associated with the wettability of the surface~\cite{vinogradova.oi:1999}. For poorly wetted hydrophilic surfaces $b=0$, i.e. the no-slip boundary condition is hold.  The slip length of hydrophobic surfaces can be of the order of tens of nanometers~\cite{charlaix.e:2005,vinogradova.oi:2003,joly.l:2006,vinogradova.oi:2009}, but not much more.

Theoretical estimates of the intrinsic and apparent slip length have been discussed in review articles~\cite{bocquet.l:2010,vinogradova.oi:2011}, and it  now  seems  certain  that  there  is not  one,  but many   mechanisms of hydrophobic slippage. They depend on substrate, contact angle and  dissolved   gas. However, the slip length of water has defied  complete understanding thus far, with accumulating experimental evidence for  surface charge- and curvature-dependent hydrodynamic slippage:
\begin{itemize}
  \item The very high surface charge is known to reduce the slip length~\cite{joly.l:2006a,xie.y:2020}. For a homogeneous charge distribution this effect can be observed when $\sigma \geq 200 $ mC/m$^2$ or $\ell_{GC} \leq 0.18$ nm~\cite{joly.l:2006a}, which is beyond a scope of the current review.
  \item Some authors revealed large and radius-dependent surface slippage in carbon nanotubes. Thus, it has been reported that $b$ shows an order of magnitude increase on reducing the carbon nanotube radius from 50 down to 15 nm~\cite{secchi.e:2016a}. In other words, when they become narrower, water flows faster. A concept of a radius-dependent slippage is not yet widely accepted and requires more experimental confirmation.
\end{itemize}

Here we keep the analysis at the simplest level by assuming that $b$ is independent on $\sigma$ and $\mathcal{L}$, but this is immaterial to our main thesis. Since the values of $b$ are comparable with the Debye lengths $\lambda_D$ of electrolyte solutions, and also with $\ell _{GC}$, it is natural to expect that hydrophobic slippage can significantly affect the whole spectrum of electrokinetic phenomena.

\subsection{Electro-hydrodynamic boundary conditions}

For more than a hundred years, colloid scientists have assumed that the potential-determining surface ions are immobile. However, recently it has been recognized that  this cannot be justified for a liquid-gas interface (bubbles or drops, foams) as well as for slippery hydrophobic solids~\cite{maduar.sr:2015}.
  The charges associated with such surfaces can migrate relative to liquid in response to $E$.  Such an adsorbed ion layer reacts to an electric field  and drags the fluid in the direction opposite to the main electroosmotic flow inside the channel. This enhances the shear stress at the interface.

To describe the fluid velocity $u_s = u|_{z=\mathcal{L}}$ at the hydrophobic surfaces that is generated by an applied electric field, an electro-hydrodynamic boundary condition has been formulated~\cite{maduar.sr:2015}
\begin{equation}
u_s =b \left[ - \partial _{z} u|_{z=\mathcal{L}}+\frac{2 (1-\mu )}{\ell _{GC}}\right],  \label{eq:bc_Stokes}
\end{equation}%
where $\mu $ can vary from 0  to 1. Integrating Eq.\eqref{eq:Stokes} and using \eqref{eq:bc_CC} one can reduce \eqref{eq:bc_Stokes} to~\cite{silkina.ef:2019}
\begin{equation}
u_s = - \dfrac{2 \mu b}{\ell _{GC}}  \label{eq:bc_Stokes2}
\end{equation}%
Thus, in the former case $u_s$ vanishes even when $b$ is large.  In the latter case $u_s$ attains its maximal value. Clearly, $u_s$ is constant for a CC channel, but it increases with $c_{\infty}$ in the CP case (see Table~\ref{table:CPCC}).

\begin{figure}[t]
	\begin{center}
		\includegraphics[width=0.9\columnwidth , trim=0.cm 0. 0.0cm
		0.,clip=false]{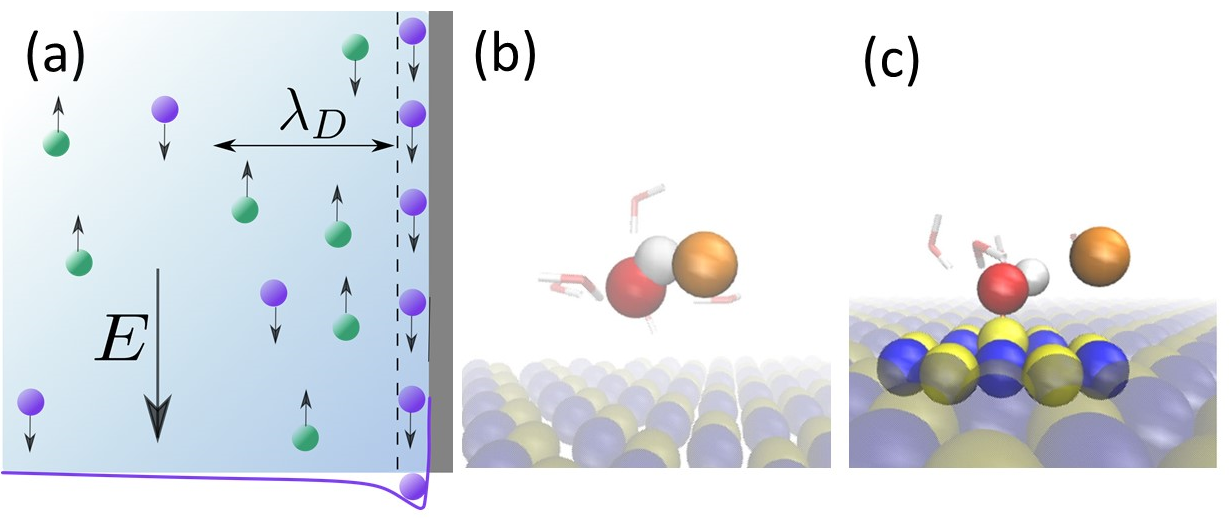}
	\end{center}
	\caption{Schematic representation of surface cations migrating under an applied electric field due to a low physisorption potential (a), and  snapshots (adapted from~\cite{mangaud.e:2022}) of chemisorbed (b) and physisorbed (c) hydroxide ions on hexagonal boron nitride surfaces.}
	\label{fig:mobile}
\end{figure}

 Returning to $\mu$, one mechanism~\cite{mouterde.t:2018} has to do with a momentum portion that (physisorbed) surface charges transfer to the wall under an applied
electric field and pressure gradient (see Fig.~\ref{fig:mobile}). Another~\cite{maduar.sr:2015} involves the ``gas cushion model''~\cite{vinogradova.oi:1995a}: the mobile surface ions (of portion $1-\mu$) are  located at the interface of a thin gas coating, but immobile ions are fixed to a solid surface itself (e.g. chemisorption). This model implies that in a pressure-driven flow, the mobile surface charges  translate with the velocity of a hydrodynamic slip given by Eq.\eqref{eq:bc_slip}, but do not migrate relative to liquid.
  Both mechanisms that are  a corollary of hydrophobicity lead to boundary condition~\eqref{eq:bc_Stokes}, but in essence, the parameter $\mu$ is still awaiting for a more detailed interpretation. The  point  is  that  such  a mobility  exists, is supported by simulation data~\cite{maduar.sr:2015,grosjean.b:2019}, but  has  immense variability depending on substrate material~\cite{mangaud.e:2022}. This subject is   becoming   more   into focus.

Finally, it should be emphasized that  Eq.\eqref{eq:bc_Stokes} becomes singular if $b=\infty$ and $\mu = 0$, which is appropriate for an interface between bulk liquid and gas. In this situation,  the boundary condition for an electro-osmotic flow takes the form~\cite{maduar.sr:2015}
\begin{equation}\label{eq:bc_Stokes3}
\partial _{z} u|_{z=\mathcal{L}} = -\frac{2}{\ell _{GC}}
\end{equation}%
The implications of condition  \eqref{eq:bc_Stokes3} for electrokinetic transport remain
largely unexplored and warrant more investigations.

\section{Mobility matrix}\label{sec:MM}

The linearity of Eq.\eqref{eq:Stokes} implies that the transport of
water and ions through the channel can be expressed in
terms of a mobility matrix $\mathcal{M}$
\begin{equation}
\begin{pmatrix}
\mathcal{U} \\
\mathcal{J}%
\end{pmatrix}%
=%
\begin{pmatrix}
M_{h} & M_{e} \\
M_{e} & K%
\end{pmatrix}%
\begin{pmatrix}
-\partial _{x}p \\
E%
\end{pmatrix}
= \mathcal{M}
\begin{pmatrix}
-\partial _{x}p \\
E%
\end{pmatrix}
\label{eq:Mo}
\end{equation}%
Here $\mathcal{U}$ [m/s] is the mean fluid velocity and $\mathcal{J}=\overline{J}+\alpha J_{\sigma}/\mathcal{L}$ [A/m$^2$] is the mean current density, where $\overline{J}$ denotes a volume-averaged value and the second term reflects the contribution of adsorbed mobile ions. The elements of matrix $\mathcal{M}$ represent so-called transport coefficients. Namely, $M_{h}$ [m$^2$/(s$\times$Pa)] is the hydrodynamic mobility, $M_{e}$ [m$^2$/(s$\times$V)] is the electroosmotic mobility, and $K$ [S/m] is the mean conductivity.

A $2\times2$ mobility matrix $\mathcal{M}$ is positive definite and symmetric (with equal off-diagonal coefficients), as assumed in \eqref{eq:Mo}, by analogy with Onsager's relations in (bulk) non-equilibrium thermodynamics~\cite{onsager.l:1931}. The equality of off-diagonal elements of matrix $\mathcal{M}$ has been confirmed for hydrophobic surfaces with immobile surface ions ($\mu=1, J_{\sigma} = 0$)~\cite{green.y:2021}, and later for the case of mobile surface charges,
providing the additional proof of validity of electro-dynamic boundary condition \eqref{eq:bc_Stokes}~\cite{vinogradova.oi:2022}. However, for a pressure induced ionic current Eq.\eqref{eq:bc_Stokes} should necessarily be supplemented by a condition~\cite{vinogradova.oi:2022}
\begin{equation}\label{eq:Js}
  J_{\sigma}=(1-\mu)\sigma U|_{z=\mathcal{L}}
\end{equation}
In other words, whatever the physical mechanism of adsorbed ion mobility is,  boundary condition \eqref{eq:bc_Stokes} would be consistent  with \eqref{eq:Mo}, if and only if, their contribution to a streaming current density is given by \eqref{eq:Js}.
A corollary of a symmetry is the so-called electro-hydrodynamic coupling
\begin{equation}\label{eq:coupling}
M_e = \dfrac{\mathcal{J}}{-\partial_{x} p}|_{E=0} \equiv \dfrac{\mathcal{U}}{E}|_{-\partial_{z} p=0}
\end{equation}
Eq.\eqref{eq:coupling} indicates that the magnitude of the streaming current induced by a pressure gradient is
directly related to that of the electro-osmotic flow generated by the applied electric field.

\begin{figure}[t]
	\begin{center}
		\includegraphics[width=0.7\columnwidth , trim=0.cm 0. 0.0cm 0.,clip=false]{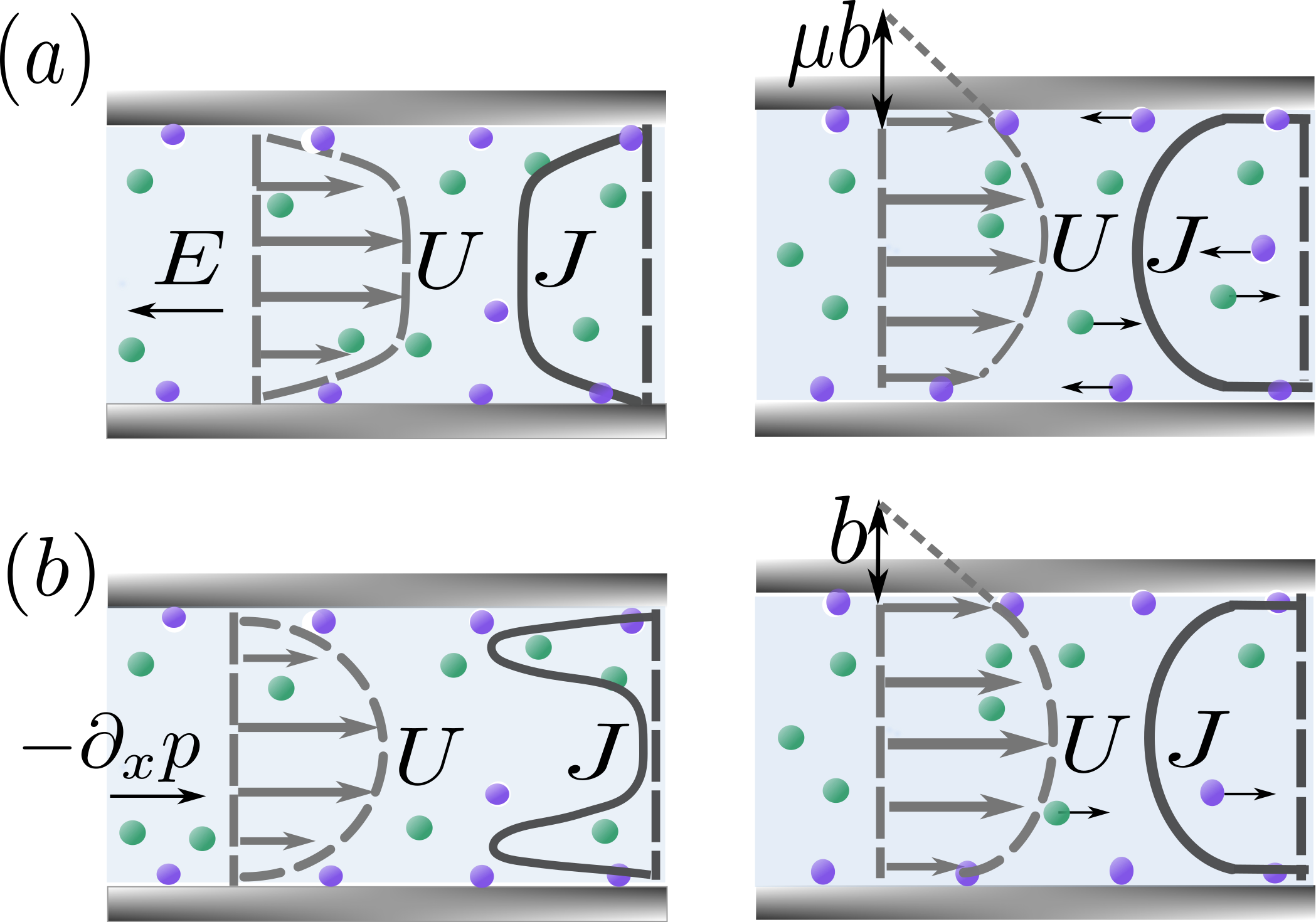}
	\end{center}
	\caption{Sketch of  $U(z)$ and $J(z)$ for channels with $\phi$ profiles as in Fig.\ref{fig:sketch}(b) for situations of $-\partial_{x} p = 0$ (a) and $E=0$ (b). The thick channel (left) is hydrophilic, whilst the thin (right) is hydrophobic.}
	\label{fig:profiles}
\end{figure}

The mobility matrix fully characterizes electrokinetic phenomena in the channel and, once its elements are known, Eq.\eqref{eq:Mo} can be used to find, without tedious calculations, the liquid flows and currents that are generated by any combination of two applied forces. We return to that in Sec.~\ref{sec:app}.
The coefficients of $\mathcal{M}$ can be obtained by setting one of two possible
driving forces to zero. Figure~\ref{fig:profiles}  shows a sketch of $U$ and $J$ profiles for these two situations discussed in detail below.

\subsection{Electro-osmotic flow and conductivity current}

If $\partial _{z}p =0$, one can reduce Eq.~\eqref{eq:Mo} to
\begin{equation}
\begin{pmatrix}
\mathcal{U} \\
\mathcal{J}%
\end{pmatrix}%
= E \begin{pmatrix}
M_{e} \\
K%
\end{pmatrix}%
\label{eq:conduction}
\end{equation}

Electro-osmotic mobility $M_{e}$ is given by
\begin{equation}\label{eq:Me}
	M_{e} = - \dfrac{e}{4\pi\eta\ell_{B}} \zeta,
	\end{equation}
where
\begin{equation}\label{eq:zeta}
	\zeta = \dfrac{e Z}{k_{B}T} = \phi _{s}-\overline{\phi } + \dfrac{2 \mu b}{\ell _{GC}} = \phi _{s}-\overline{\phi } - u_s
\end{equation}
is the dimensionless zeta (or
electrokinetic) potential.
Thus, zeta potential of hydrophobic channels includes a contribution of an electrostatic potential as well as a slip velocity at the wall, which in turn depends both on electrostatic (surface charge) and  wetting (hydrophobic slip length and surface charge mobility) properties of the walls.

When $\mathcal{L} \to \infty$, which is the case of an isolated surface, $\overline{\phi } \to 0$, a relation between $\phi_s$ and surface charge density is given by Eq.\eqref{eq:pot-charge_hs}, and $\zeta$ becomes a property of the surface itself. However, zeta potential of hydrophobic surfaces
no longer reflects the sole surface potential, except the situation of $\mu=0$, since it is enhanced due to a finite fluid velocity at the walls
\begin{equation}\label{eq:zeta_single}
  \zeta = \phi_s - u_s
\end{equation}
(or $Z = \Phi_s + \dfrac{4 \pi}{\varepsilon} \mu b \sigma$). A sketch of $U(z)$ near a slippery wall is included in Fig.~\ref{fig:smol} along with the Smoluchowski profile.
As a side note, similar mechanism of zeta potential enhancement is observed for hydrophilic surfaces coated by porous nanofilms that are permeable for water and ions~\cite{vinogradova.oi:2020,silkina.ef:2021}. For a homogeneous hydrophilic surface ($b=0$), however, $\zeta = \phi_s$, which is the classical Smoluchowski result~\cite{smoluchowski.m:1921}.

It is instructive now to clarify how to use Eq.\eqref{eq:zeta_single} in the CC and CP cases. Substituting \eqref{eq:pot-charge_hs} we obtain the CC expression
\begin{equation}\label{eq:zeta_CC}
  \zeta = 2\arsinh\left(\frac{\lambda _{D}}{\ell _{GC}}\right) + \dfrac{2 \mu b}{\ell _{GC}},
\end{equation}
which, say, for a low effective surface charge,  $\lambda_{D}/\ell _{GC}\leq 1$, gives $\phi_s \simeq 2 \lambda _{D}/\ell _{GC}$ and $\zeta/\phi_s \simeq 1 + \mu b/\lambda_{D}$. Thus, although $\phi_s$ itself decreases with salt, in concentrated solutions $\zeta \gg \phi_s$, provided $\mu b/\lambda_{D}$ is large. By expressing $\ell _{GC}$ in \eqref{eq:zeta_CC} through $\phi_s$ (see Table \ref{table:CPCC}) one can immediately obtain a solution for the CP case
\begin{equation}\label{eq:zeta_CP}
  \zeta = \phi_s + \dfrac{2 \mu b}{\lambda_{D} }\sinh \dfrac{\phi_s}{2}
\end{equation}
For $\phi_s \leq 1$ this gives the same $\zeta/\phi_s$ as in the CC case, but since the surface potential is constant, $\zeta$ becomes very large at high salt. Thus, if the hydrophobic slippage is ignored, one can erroneously infer a huge $\phi_s$ from the measurements.

\begin{figure}[t]
\begin{center}
\includegraphics[width=0.99\columnwidth , trim=0.cm 0. 0.0cm
0.,clip=false]{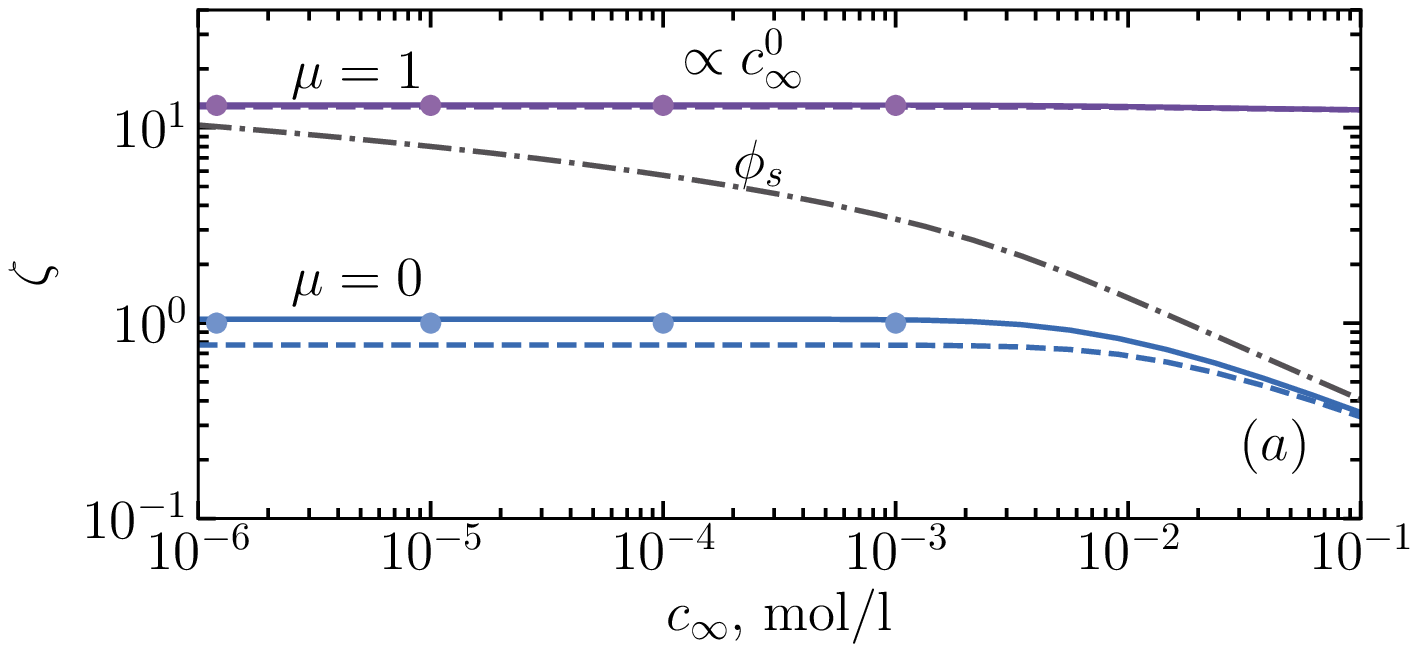}
\includegraphics[width=0.99\columnwidth , trim=0.cm 0. 0.0cm
0.,clip=false]{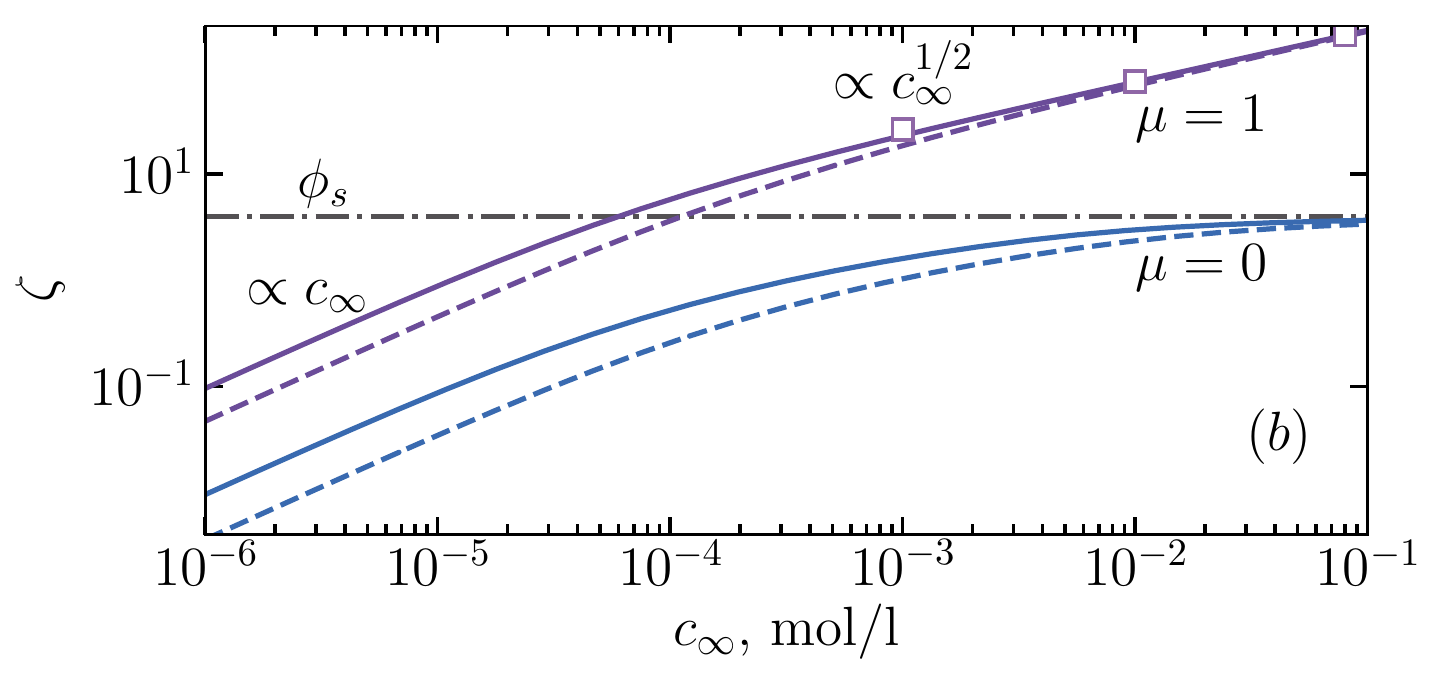}
\end{center}
\caption{Zeta potential as a function of salt concentration computed for a channel of $\mathcal{L} = 10$ nm and $b = 30$ nm using fixed $\ell_{GC}=5$ (a) or $\phi_{s}=4$ (b). Solid curves represent $\zeta$ of a slit, dashed ones - of the cylinder. The surface potential computed for a cylinder is shown by dash-dotted curves. Squares and circles show calculations from Eqs.\eqref{eq:zeta_CP} and \eqref{eq:av_pot} with $\alpha = 2$.  }
\label{fig:zeta}
\end{figure}

In the general case, the zeta potential is a global electrodynamic property of the channel. Finite  $\overline{\phi }$ reduces its value, but the  hydrophobic slippage if any tends to  augment $\zeta$. The competition between these effects defines a  magnitude  of $\zeta$, which can be larger or smaller than $\phi_s$, and provides a diversity of its behavior. This is illustrated in Fig.~\ref{fig:zeta}, where $\phi_s$ is compared with the values of $\zeta$ computed using $\mu = 0$ and 1 (both for a slit and a cylinder of $\mathcal{L} = 10$ nm and $b = 30$ nm). It can be seen that there exists some quantitative difference between cylinders and slits, but the qualitative features of the $\zeta$-curves are the same. The CC calculations shown in Fig.~\ref{fig:zeta}(a) are made with the value of $\ell _{GC} = 5$ nm.
In the case of $\mu=0$, identical to a no-slip channel, $\zeta = \phi_s - \overline{\phi } \leq \phi_s$. Zeta potential  remains constant at $c_{\infty} \leq 10^{-3}$ mol/l (which is the thin channel mode, as discussed in Sec.~\ref{sec:BCCT}) and begins to converge slowly to $\phi_s$ only at a higher concentration. The emergence of the plateau, which points out that $\mathcal{U}$ becomes independent on $c_{\infty}$, is an entirely unexpected result that is impossible for a single hydrophilic wall.
When $\mu = 1$, $\zeta$ is large, exceeds $\phi_s$, and depends neither on amount of salt nor geometry. In the thin channel mode the difference between $\phi_s$ and $\overline{\phi}$ can be calculated analytically, so is $\zeta$
\begin{equation}\label{eq:av_pot}
 \phi_s - \overline{\phi}\simeq \dfrac{2}{2+\alpha} \dfrac{\mathcal{L}}{\ell _{GC}}, \, \zeta \simeq -u_s + \dfrac{2}{2+\alpha} \dfrac{\mathcal{L}}{\ell _{GC}}
\end{equation}
Consequently, when $\mu b/\mathcal{L} \ll 1$, one can derive $\zeta \propto \mathcal{L}/\ell _{GC}$. This is the case of $\mu = 0$ in Fig.~\ref{fig:zeta}(a): the value of $\zeta$ at saturation reflects only (constant) $\sigma$. However, if $\mu b/\mathcal{L} \gg 1$, then $\zeta \simeq -u_s\propto \mu b/\ell _{GC}$, so that when $\mu =1$, the plateau appears due to a constant $b \sigma$ and can be used to infer its value. In the thick channel mode it is generally assumed that $\overline{\phi } \simeq 0$, so the Eq.\eqref{eq:zeta_single} could be sensible approximation. The same calculations, but for the CP channel of $\phi_s = 4$ are shown in Fig.~\ref{fig:zeta}(b). One important conclusion is that $\zeta$ now grows with $c_{\infty}$. If $\mu = 1$, the zeta potential significantly exceeds $\phi_s$ in the thick channel mode, and for this
branch of the curve $\zeta \propto c_{\infty}^{1/2}$.
Also included in Fig.~\ref{fig:zeta} some analytical results obtained for CC and CP cylinders (that constitute a more realistic model for artificial nanotubes and real porous materials). It can be seen that the fit is quite good.

These results imply that if $\mu b = 0$ any CR model should lead to a power-law scaling $\zeta \propto c_{\infty}^{\gamma}$) with an exponent between 0 and 1 in the thin channel mode, or  from -1/2 to 0 in the thick channel mode. A finite $\mu b$ not only increases $\zeta$,  but also leads to a different scaling: in concentrated solutions $\gamma$ is now bounded by 0 and 1/2.

Some further  comments   should   be  made. That $\overline{\phi }$  augments on dilution is apparent since
EDLs begin to occupy a large portion of the channel. In essence, the emergence of the plateau branch only indicates that $\phi_s - \overline{\phi } \propto c_{\infty}^0$, but it does not require a thin channel mode or even an overlap of EDLs. The analytical calculations  of $\phi_s - \overline{\phi }$ beyond the thin channel mode appear to be missing  and remain a challenge.

\begin{figure}[t]
\begin{center}
\includegraphics[width=0.99\columnwidth , trim=0.cm 0. 0.0cm
0.,clip=false]{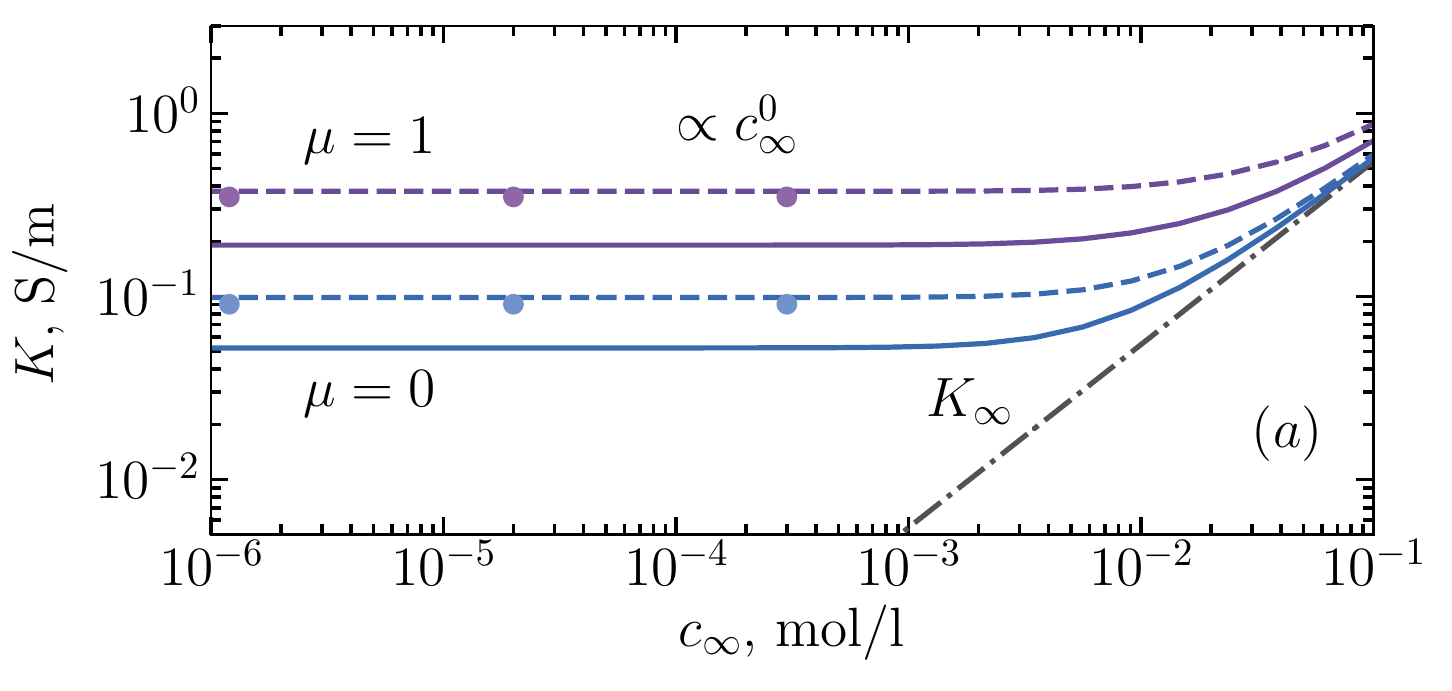}
\includegraphics[width=0.99\columnwidth , trim=0.cm 0. 0.0cm
0.,clip=false]{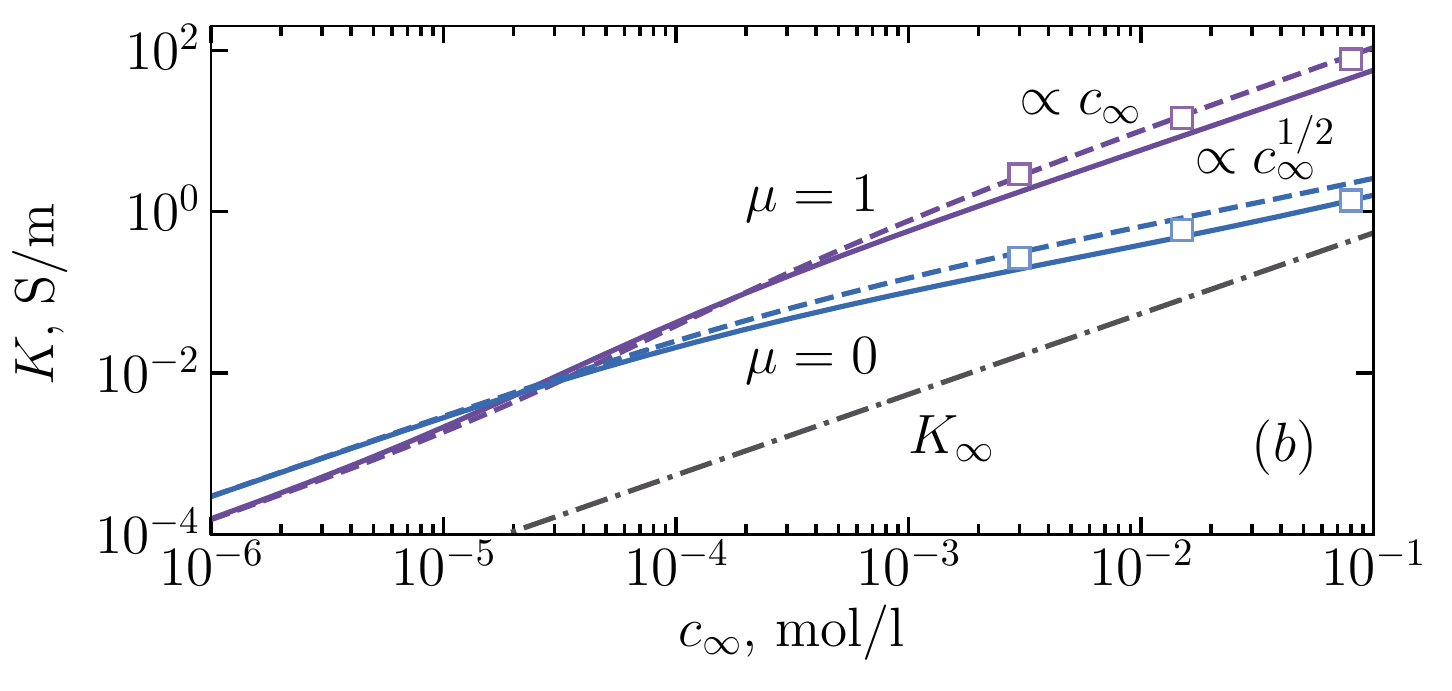}
\end{center}
\caption{Electrical conductivity as a function of salt concentration computed for the same CC and CP channels as in Fig.~\ref{fig:zeta}.
Dash-dotted lines show conductivity of the bulk electrolyte. }
\label{fig:cond}
\end{figure}

Finally, we turn to the mean conductivity of the channel $K$. Numerical results are shown in Fig.~\ref{fig:cond} and compared with $K_{\infty}$. The calculations
are made for the same channels as in Fig.~\ref{fig:zeta}. One can see an obvious correlation   with   zeta potential data:  in the CC case the conductivity plateau appears, and exactly in the same concentration range. The conductivity amplification at the plateau branch compared to the bulk is huge, several order of magnitudes, and depends on $\mu$ (and geometry).

It is conventional to divide $K$ into two contributions: $K_0$ arising for hydrophilic channel, and slip-driven $\Delta K$ that is associated with hydrophobicity
\begin{equation}
K = K_0 + \Delta K
\label{eq:M22}\end{equation}

For a channel of any thickness
\begin{equation}\label{eq:K0}
K_0 = K_{\infty }\displaystyle \left[ \dfrac{3 \lambda _{D}^{2}\mathcal{R}\overline{(\phi^{\prime} )^{2}}}{2\ell _{B}}+ \overline{\cosh {\phi }}\right],
\end{equation}
where $K_{\infty }$ is given by Eq.\eqref{eq:ic0_thick}
The first and second terms in \eqref{eq:K0} are associated with the convective and migration contributions, correspondingly.
To use \eqref{eq:K0} the mean square derivative
of the electrostatic potential $\overline{(\phi^{\prime})^2}$, which is the
measure of the electrostatic field energy (per unit area), and the mean osmotic pressure $\overline{\cosh\phi}$
 should be substituted. Their detailed calculations for slits and cylinders can be found in \cite{vinogradova.oi:2021,vinogradova.oi:2022} and the results for $K_0$ can be summarized as follows. The surface conductivity, i.e. the conductivity associated with the EDL, dominates over the bulk one when
$\ell _{Du}/\mathcal{L}=O(1)$ or larger. If so,
\begin{equation}\label{eq:K0_CC}
K_0 \simeq K_{\infty }\dfrac{2 \alpha \ell _{Du}}{\mathcal{L} }\beta,
\end{equation}
with $\beta=O(1)$. Namely, it is equal to $1+3 \mathcal{R}/\ell _{B} \simeq 2$ in the thick and $\simeq 1$ in the thin channel mode. This implies that in CC channels $K_0 \propto e\alpha\beta\sigma/\mathcal{L}$ does not depend on the salt concentration (since $K_{\infty } \ell _{Du} \propto c_{\infty}^0$). Thus, the surface contribution to the conductivity shows up as a saturation of the conductivity in dilute solutions, where the bulk contribution is practically absent~\cite{stein.d:2004,schoch.rb:2005}. The height of the conductivity plateau augments on increasing surface charge density and reduce on decreasing $\mathcal{L}$ (but the conductance $\propto K_0 \mathcal{L}$ does not depend on the channel thickness). By expressing $\ell _{Du}$ in \eqref{eq:K0_CC} through $\phi_s$ one can obtain $K_0$ in the CP case. In the thick channel mode
\begin{equation}\label{eq:K0_CP_thick}
K_0  \simeq K_{\infty }\dfrac{2 \alpha \lambda_D}{\mathcal{L} } \sinh \dfrac{\phi _{s}}{2} \propto c_{\infty}^{1/2},
\end{equation}
but for a thin channel mode $K_0 \propto  K_{\infty}  \sinh \phi_s \propto c_{\infty}$.

The slip-driven contribution $\Delta K$ is given by
\begin{equation}\label{eq:slip correction}
\Delta K =  K_{\infty }\frac{2 \alpha \ell_{Du}}{\mathcal{L}} \left(1-\mu - \frac{3 \mu u_s}{2}\dfrac{\mathcal{R}}{\ell _{B}} \right),
\end{equation}
where the proportional to $(1 - \mu)$ term is associated with a migration contribution of adsorbed ions and the last term  accounts for an additional convective conductivity due to a finite slip velocity. Eq.~\eqref{eq:slip correction} was first derived for a thick slit~\cite{mouterde.t:2018}, but a later study has proven that it is valid for any channel~\cite{vinogradova.oi:2021}.

It follows then from \eqref{eq:M22} that
\begin{equation}\label{eq:K1_CC}
K \simeq K_{0}\left(1+ \dfrac{1-\mu}{\beta} + \frac{3 \mu^2 b}{\beta \ell _{GC}}\dfrac{\mathcal{R}}{\ell _{B}} \right),
\end{equation}
where the expression in the brackets represents the amplification of conductivity due to hydrophobization. It is nonlinear function of $\mu$, which can exhibit a minimum~\cite{vinogradova.oi:2021}. For the sake of brevity here we only mention the cases of $\mu=0$, where only a migration of surface ions generates a complimentary current, and of
$\mu=1$, where only a convective ionic current enhances the conductivity. In the former case $K/K_0 \simeq 1 + 1/\beta$, i.e. the conductivity increases, and can be twice larger than $K_0$ (in thin channel mode). Importantly, in any modes, and independently of electrostatic boundary conditions,  $K$ at $\mu=0$ scales with $c_{\infty}$ exactly as $K_0$ does.
In the case of $\mu=1$
\begin{equation}\label{eq:ampK1}
  K \simeq K_0 \left(1+\frac{3 b}{\beta \ell_{GC}}\dfrac{\mathcal{R}}{\ell _{B}}\right)
\end{equation}
The amplification of conductivity, $K/K_0 \propto b/\ell_{GC}$,
can be huge, provided $b/\ell_{GC}$ is large. If so, in the CC case $K$ does not depend on salt and the effect of slippage shows up simply in the shift of the saturation plateau towards a much larger value.
If we deal with the CP channel of $\mu = 1$, the scaling of $K$ with salt becomes different. It is straightforward to show that in the thick channel mode
\begin{equation}
  K \simeq K_0 \left( 1+\frac{ b }{ \lambda_D}\dfrac{3 \mathcal{R}}{\beta \ell _{B}}\sinh \dfrac{\phi _{s}}{2}\right)
\end{equation}
If the second term dominates, $K \propto c_{\infty}$, and slippage appears as a shift of the bulk conductivity curve.
In the thin channel mode
\begin{equation}
  K \simeq K_0 \left( 1+\dfrac{b \mathcal{L} }{ 2\alpha \lambda_D^2}\dfrac{3 \mathcal{R}}{\beta\ell _{B}}\sinh \phi _{s}\right),
\end{equation}
indicating that the  conductivity increment does exist and is salt-dependent, but quite small. Thus, we might argue that a sensible scaling should be $K \propto c_{\infty}$, but note that conductivity can become smaller than at $\mu = 0$. At first sight this is surprising, but we recall that in very dilute solutions $K$ is twice larger than $K_0$. Some of these theoretical results are included in Fig.~\ref{fig:cond}. It can be seen  that the approximate formulas are
very reliable.

Summarising the scaling properties of the CC and CP channels, one can expect that any CR model should lead to $K_0 \propto c_{\infty}^{\gamma}$ with $\gamma$ bounded by 0 and 1 in the thin channel mode or 0 and 1/2 in the thick channel mode. The values of $\gamma = 1/3$~\cite{secchi.e:2016} and 1/2~\cite{biesheuvel.pm:2016} obtained for dilute solutions lie in between these attainable bounds. These two results have been actively discussed in the
literature~\cite{uematsu.y:2018,manghi.m:2018,green.y:2022}, but in essence, in the thin channel mode, $\gamma$ can vary smoothly from 0 to 1 depending on parameters of the CR model. Hydrophobic slippage has the effect of allowing augmented conductivity, $ K\propto c_{\infty}^{\gamma}$, but if $\mu = 0$ the scaling behavior does not change. For $\mu = 1$ in the thin channel mode $0 \leq \gamma \lesssim 1$, and in the thick channel mode $1/2 \leq \gamma \leq 1$. Thus, the hydrophobicity not just increases the conductivity, but augment $\gamma$, although  in concentrated solutions only. To the best of our knowledge such predictions have not been tested in experiment and simulations yet.

\subsection{Pressure-driven flow and streaming current}

If $E = 0$, Eq.\eqref{eq:Mo} reduces to
\begin{equation}
\begin{pmatrix}
\mathcal{U} \\
\mathcal{J}%
\end{pmatrix}%
= -\partial _{z}p \begin{pmatrix}
M_{h} \\
M_{e}%
\end{pmatrix}%
\label{eq:streaming}
\end{equation}

Thus, the streaming current measured as a function of applied pressure gradient allows one to determine $M_e$, and then to use Eq.\eqref{eq:Me} to infer $\zeta$.
From  a   pragmatic  view,  the  choice of this experiment  is dictated by the complexity (or  impossibility)  of the
fluid velocity measurements in narrow channels. In any event, the streaming current studies are always much easier to perform than any velocimetry experiment.

The hydrodynamic mobility, or a coefficient which relates applied pressure gradient with the mean velocity is given by
\begin{equation}\label{eq:Mh}
M_{h} = \dfrac{\mathcal{L}^2}{\alpha (2+\alpha) \eta} \left( 1 + \dfrac{(2 + \alpha) b}{\mathcal{L}}\right)
\end{equation}
The value of $M_{h}$ thus depends on the size and geometry of the channel and the wetting properties of its walls, but does not depend on electrostatics.

\section{Applications}\label{sec:app}

The main purpose of Sec.~\ref{sec:MM} has been to
show that  electrokinetic phenomena in a narrow channel can be significantly different from those near a single wall, and also why and when. This  may
open many novel applications. We list below what we believe are the most relevant.

\subsection{Probing surface properties}

 By measuring (streaming or conductivity) currents and plotting $\zeta$ and/or $K$ against $c_{\infty}$, the surface properties
can be tested. That the surface is of the constant $\sigma$ is signalled by an emergence in dilute solutions of the zeta potential and conductivity  plateau. The height of the conductivity plateau was long time used to infer $\sigma$ of hydrophilic silica, but the amount of its shift due to hydrophobicity reflects the values of   $\mu$ and $b$ that can also be determined. Such a procedure is beset with difficulties. The point is that the electrostatic
and hydrodynamic effects are strongly coupled. To disentangle them, a series of several different (multi-step) experiments is required. Again,  evidences of large $\zeta$ and $K$ in concentrated solutions, coupled with some particular exponents of power-law scaling are signalling   that surfaces keep $\phi_s$ constant (or obeying CR conditions) and   point to the absence or existence of slip and surface charge mobility. A properly designed experiment would allow to infer their magnitude.

\subsection{Energy conversion}
The amplification by hydrophobic slippage of all elements of the mobility matrix $\mathcal{M}$  opens very interesting
perspectives in the context of electrokinetic energy conversion.
\begin{itemize}
  \item The emergence of streaming current represents a tool for conversion
hydrostatic energy into electrical power. The maximum efficiency, i.e. the
ratio of the output to the input power is~\cite{vanderHeyden.fhj:2006}%
\begin{equation}
\mathcal{E}=\frac{\Theta}{\Theta +2\left( \sqrt{1-\Theta }+1-\Theta
\right) },
\end{equation}%
where $\Theta =M_{e}^{2}/\left( KM_{h}\right).$ Slip lengths of a few tens of nanometers are predicted to increase the
efficiency of the energy conversion up to 40\%~\cite{ren.y:2008}. However, the mobility of surface charges significantly reduces $E$~\cite{Liu.Y:2022}. These results would require  thorough experimental validation with various hydrophobic materials.

 \item The electrokinetic energy can be converted to a mechanical energy. For example, one can generate a high pressure gradient at the nanoscale. Suppose we close one end of the channel and apply $E$. Substituting $\mathcal{U}=0$ to \eqref{eq:Mo} leads to $ -\partial _{z}p/E = - M_{e}/M_{h}$. Could the hydrophobicity augment $-\partial _{z}p/E$? The answer to this question is by no means obvious since both $M_e$ and $M_h$ increase with $b$. Using \eqref{eq:av_pot} and \eqref{eq:Mh} one can be readily demonstrated that  for the nanotube of  $\mathcal{L} = 10$ nm and $b=30$ nm in the thin channel regime $-\partial _{z}p/E$ ``increases'' in $\mu$ times. Thus, a hydrophobicity has detrimental or no effect. However, in the thick channel regime $ \partial _{z}p/E$ for a given hydrophobic cylinder augments in ca. $\mu \mathcal{L}/4 \lambda_D$ times. This implies that for $\mu=1$ and $c_{\infty} = 10^{-1}$ mol/l the amplification of $ \partial _{z}p/E$ in 2.5 times can be expected. To our knowledge such predictions have not been made before, nor tested in experiment.

\end{itemize}

\section{Conclusion}

This  review  was  motivated  by  an  awareness  of many unusual experimental results, in part on  giant conductivities and zeta potentials measured in micro- and nanochannels.  These  observations  have posed serious  issues for classical  theories  of  electrokinetic phenomena, and also about the manifestations of hydrophobicity. The effects of hydrophobic slippage until recently have not  been considered, and the new ``ingredient'' is the existence of a migration of adsorbed surface ions with respect to liquid under an electric field. During the last several years theory has made striking  advances leading to interpretation of electrokinetic experiments in hydrophilic channels, as well as predictions of novel effects that might be expected to occur when channels are hydrophobic. The time is probably right for more detailed simulations and experimental studies. If these effects were better clarified and tested, the implications are large.

\section*{Acknowledgements}
This work was supported by the Ministry of Science and Higher Education of the Russian Federation.


%
%
%
%
%
%


\bibliographystyle{unsrt}
\bibliography{cocref}

\end{document}